\documentclass[aip,pop,reprint]{revtex4-1}

\usepackage{hyperref}
\usepackage[utf8]{inputenc}
\usepackage[english]{babel}
\usepackage{bm}
\usepackage{amsmath}
\usepackage[shortcuts]{extdash}
\usepackage{graphicx}
\usepackage{tabularx}
\usepackage{soul,xcolor}
\DeclareUnicodeCharacter{2010}{-}

\begin{document}
\DeclareGraphicsExtensions{.pdf,.png,.jpg}

\title{Terahertz emission from submicron solid targets irradiated by ultraintense femtosecond laser pulses}

\author{J. D\'echard}
\author{X. Davoine}
\author {L. Gremillet}
\email{laurent.gremillet@cea.fr}
\author{L. Berg\'e}
\email{luc.berge@cea.fr}
\affiliation{CEA, DAM, DIF, F-91297 Arpajon, France}
\affiliation{Universit\'e Paris-Saclay, CEA, LMCE, 91680 Bruy\`eres-le-Ch\^atel, France}

\date{\today}

\begin{abstract}
Using high-resolution, two-dimensional particle-in-cell simulations, we investigate numerically the mechanisms of terahertz (THz) emissions in submicron-thick carbon solid foils driven by ultraintense ($\sim 10^{20}\,\rm W\,cm^{-2}$), ultrashort ($30\,\rm fs$) laser pulses at normal incidence. The considered range of target thicknesses extends down to the relativistic transparency regime that is known to optimize ion acceleration by femtosecond laser pulses. By disentangling the fields emitted by longitudinal and transverse currents, our analysis reveals that, within the first picosecond after the interaction, THz emission occurs in bursts as a result of coherent transition radiation by the recirculating hot electrons and antenna-type emission by the shielding electron currents traveling along the fast-expanding target surfaces.
\end{abstract}

\maketitle

\section{Introduction} 

Terahertz (THz) waves, lying in between microwaves and optical waves, are of growing interest in various areas of research and industry covering
medical imaging, remote detection, time-resolved molecular spectroscopy, cryptography, telecommunications, cultural heritage or environment
\cite{Tonouchi:np:2007, Berge:epl:2019}. Nowadays, the operational bandwidth of THz sources is not restricted to the ``THz gap'',
but is routinely extended up to mid-infrared frequencies.
Recently, novel challenges such as the development of compact THz electron accelerators
\cite{Nanni:nc:2015, Sharma:jpb:2018,Curcio:sr:2018} or THz-triggered chemistry \cite{LaRue:prl:2015} raised the need of mJ energy THz pulses
with field strength at the $\rm GV\,m^{-1}$ level. As of now, optical rectification \cite{Vicario:prl:2014} or the tilted-pulse-front technique
\cite{Fulop:oe:2014} in nonlinear crystals can achieve sub-mJ THz energies with a few $0.1\,\rm GV\,m^{-1}$ field strengths, yet further progress
in these solid-based technologies is hampered by their inherent damage threshold. By contrast, plasma spots created by intense laser pulses may
supply suitable damage-free emitters \cite{Hafez:jo:2016}. Such is the case with gases ionized by  two-color laser pulses (e.g., fundamental and
second harmonics) of moderate intensities $\sim 10^{14}\,\rm W\,cm^{-2}$,  which, through the generation of photocurrents, have been shown to radiate
ultrabroadband ($\geq 100\,\rm THz$), relatively strong ($\geq 0.1$ GV/m) terahertz pulses \cite{Kim:np:2008}.
Using a $3.9\,\rm \mu m$ pump laser wavelength, this mechanim lately resulted in $\lesssim 0.2\,\rm mJ$ THz pulses, associated with $\sim 1\,\%$
laser-to-THz conversion efficiency \cite{Nguyen:pra:2018,Jang:optica:2019,Koulouklidis:nc:2020}. However, when applied to standard $\sim~1\,\rm \mu m$
laser wavelengths, it seems to be limited to the production of $\sim \rm \mu\,J$ energy THz pulses only \cite{Oh:apl:2013}.

Higher THz energy yields, approaching the mJ level, may be attained in relativistic gas-jet interactions \cite{Dechard:prl:2018, *Dechard:prl:2019} through coherent transition radiation (CTR) by wakefield-accelerated electrons \cite{Leemans:prl:2003, *Schroeder:pre:2004}. This radiation is coherent in the THz frequency domain as the corresponding wavelengths exceed the typical dimensions of the fast electron source; its power therefore scales quadratically with the number of fast electrons exiting the plasma. The promising potential of this wakefield accelerator-based setup, particularly in the matched-blowout regime as predicted by particle-in-cell (PIC) numerical simulations \cite{Dechard:prl:2018, *Dechard:prl:2019}, still awaits experimental  demonstration. On the other hand, the capability of relativistic laser-solid interactions to give rise to few $10\,\rm mJ$ THz pulses has already been evidenced~\cite{Liao:pnas:2019}. The available data suggest that THz emission then proceeds through a variety of mechanisms, all triggered by the motion of laser-driven high-energy  electrons.

When an intense laser pulse impinges onto a solid target, the latter is rapidly ionized and a population of energetic electrons is generated via various processes (e.g., $J\times B$ heating~\cite{Kruer:pof:1985, Debayle:pop:2013}, vacuum heating~\cite{Brunel:prl:1987, Bauer:pop:2007, May:pre:2011} or resonant absorption~\cite{Forslund:pra:1975, Hu:pop:2006}), depending on the laser intensity, incidence angle and density scale length~\cite{Kemp:nf:2014}. The dynamics of those hot electrons, together with the collective response of the target bulk electrons and ions, can initiate a number of radiative processes in the THz frequency domain. This is so because, as discussed below, their characteristic time scales are determined by either the laser duration, the light transit time through the target, or the ion acceleration time, all ranging from a few $10\,\rm fs$ to $\sim 1\,\rm ps$.

Let us now consider the case of steep-gradient, micron-thick targets, irradiated at normal incidence by a relativistically intense ($\gtrsim 10^{18}\,\rm W\,cm^{-2}$) short-pulse laser. Firstly, the crossing of the target surfaces by the longitudinally accelerated fast electrons generates CTR over broad frequencies \cite{Zheng:pop:2003, Schroeder:pre:2004, Bellei:ppcf:2012}. Unlike gases where CTR is highly collimated on axis due to ultrarelativistic, low-divergence fast electrons \cite{Dechard:prl:2018},  CTR from solid-targets seems to be emitted at broader angles as a result of less energetic and more divergent electrons \cite{Ding:pre:2016}. This radiation is accompanied by an outward transverse current pulse that travels close to the speed of light along the target surface, and acts to screen the fields inside the plasma \cite{Tokita:sr:2015}. This transient surface current radiates like the charge image of the energetic electron bunch, and is an intrinsic feature of transition radiation \cite{Ginzburg:ps:1982}. 

Now, very few electrons are able to escape the potential barrier set up around the positively charged target, most of them being drawn back into the target \cite{Myatt:pop:2007, Poye:pre:2015}. If the target is thicker than the longitudinal extent of the recirculating hot electrons, yet much shorter than their (millimeter-range) deceleration length, their back and forth motion across the target induces periodic CTR bursts \cite{Ding:pre:2016}. When exiting the target, those electrons can propagate a distance of the order of their Debye length ($\sim \rm \mu m$) before being reflected. There ensue hot-electron sheaths at the target surfaces, which expand radially at a velocity $\lesssim c$ due to some electrons accelerated by the laser at large angles from the target normal or deflected by self-generated fields. Note that, at oblique laser incidence, the sheath formed at the irradiated side may comprise hot electrons flowing along the target surface due to counteracting self-induced electric and magnetic fields \cite{Nakamura:prl:2004}. THz emissions can also take place during the emergence phase of those sheaths.

The quasistatic fields generated by the outward-moving hot-electron sheaths make up another inductive source of shielding surface currents. In PIC simulations of laser-solid interactions, owing to the relatively small ($\lesssim 100\,\rm \mu m$) target sizes generally considered, the above two types of surface current pulses, which respond, respectively, to the CTR by the hot electrons crossing the target boundaries and the static fields due to the laterally moving hot electrons, are difficult to disentangle because of their relativistic velocities. In the following, we will refer to them under the generic term of ``shielding (surface) currents''. Whereas they do not radiate while flowing steadily along the target surfaces, they can emit further radiation when reaching the transverse target edges \cite{Zhuo:pre:2017} in an antenna-like fashion \cite{Smith:ajp:2001}.
  
The electrostatic fields associated with the hot-electron sheaths eventually set into motion the target ions. The protons present in the target bulk or as surface contaminants respond the fastest given their large charge-to-mass ratio. This is the well-known target normal sheath acceleration (TNSA) mechanism, known to dominate ion acceleration
in micrometric foils driven at laser intensities $\lesssim 10^{20-21}\,\rm W\,cm^{-2}$~~\cite{Wagner:prl:2016}. This results in the outward expansion of a quasineutral electron-proton plasma, preceded, at its front, by a negatively charged double layer \cite{Mora:prl:2003, *Mora:pre:2005}. The relatively long-time-scale acceleration of the latter is the source of a 
wide-angle dipole-like radiation -- often termed sheath radiation (SR) \cite{Gopal:njp:2012, *Gopal:prl:2013, *Gopal:ol:2013} --, scaling quadratically with the net charge of the double layer. 

A record-high $10.5\,\rm mJ$ THz yield, corresponding to a $\sim 1.7\%$ laser-to-THz energy conversion efficiency, was achieved by Jin \emph{et al.} \citep{Jin:pre:2016}, using a $\sim 3\times 10^{19}\,\rm Wcm^{-2}$ intensity, $30\,\rm fs$ duration laser pulse and Cu foils with thicknesses  $2\le d_0 \le 30\,\rm \mu m$. The energy of the THz radiation, peaking at about $\pm 45^\circ$ from the target rear normal, was found to scale as $1/d_0^3$, consistent with the prediction of a simple SR model. An interesting feature was the development of multiple THz pulses over ps time scales, the number of which increasing in thinner targets. Lately, under similar conditions, Herzer \emph{et al.} \cite{Herzer:njp:2018} managed to discriminate between the CTR and SR contributions. The latter was observed to yield a broader  dipole-like angular distribution than the former, but with a much larger energy yield ($\sim 700\,\rm \mu J$ vs. $\sim 40\,\rm \mu J$). In contrast to Ref.~[\onlinecite{Jin:pre:2016}], the THz waveform then exhibited a single-cycle shape of $\sim 1\,\rm ps$ duration.

Past simulation studies on THz emissions from laser-solid interactions have only considered micrometer-range target thicknesses. Such targets, however, may not be the most effective neither in terms of laser-to-hot-electron coupling nor of ion acceleration. The purpose of this paper is, rather, to investigate THz radiation from submicron foils irradiated by ultraintense femtosecond laser pulses, including nanometric targets enabling relativistic self-induced transparency (RSIT) of the laser pulse~\cite{Vshivkov:pop:1998}. The threshold conditions for this regime have been shown to entail a strong coupling efficiency into hot electrons and to enhance ion acceleration \cite{dHumieres:pop:2005, Esirkepov:prl:2006, Brantov:prstab:2015, Ferri:arxiv:2020}. The optimal target thickness for ion acceleration, also corresponding to the onset of RSIT, is given by \cite{Esirkepov:prl:2006, Brantov:prstab:2015}
\begin{equation} \label{eq:d_opt}
d_{\rm opt} \simeq a_0  \frac{n_c}{2n_e}\lambda_0,
\end{equation}
where $a_0 \equiv eE_0/m_e c \omega_0$ is the dimensionless laser field strength ($E_0$ is the laser field strength, $\omega_0$ the laser frequency, $m_e$ the electron mass, $e$ the elementary charge and $c$ the speed of light in vacuum), $\lambda_0 \equiv 2\pi c/\omega_0$ the laser wavelength, $n_c \equiv m_e \omega_0^2 \epsilon_0/e^2$ the related critical density ($\epsilon_0$ is the vacuum permittivity) and $n_e$ the target electron density. For fiducial parameters ($a_0 = 10$, $n_e/n_c = 100$), $d_{\rm opt}$ is typically of a few 10~nm.

Most previous works on THz radiation from laser-solid interactions considered obliquely incident laser pulses \cite{Gopal:njp:2012, Ding:pre:2016, Jin:pre:2016, Zhuo:pre:2017, Herzer:njp:2018}, yet this setup complicates the hot-electron generation and subsequent dynamics, besides causing a natural asymmetry in the angular distribution of the THz radiation. This makes it more intricate to distinguish the various THz radiation processes at play, and their related electron current sources. Therefore, to simplify the analysis, the present study will only address the case of a normally impinging laser pulse.

This paper is structured as follows. The simulation parameters are detailed in Sec.~\ref{sec:setup}. In Sec.~\ref{sec:main_features} are reported the main features of the laser-driven  electron and ion dynamics for a reference 500-nm-thick CH$_2$ target. The resulting THz radiation emitted within the first 100~fs is analyzed in Sec.~\ref{sec:general_analysis}, notably by discriminating between the longitudinal and transverse current sources, and by untangling the contributions of the hot-electron and shielding surface currents. Several radiation processes are identified and the variations of their properties with the target thickness over a $\sim 1\,\rm ps$ time scale are addressed in Sec.~\ref{sec: foil_thickness}. Our results are summarized in Sec.~\ref{sec:conclusion}.

\section{Simulation setup}
\label{sec:setup}

Our PIC simulations have been performed using the fully relativistic, electromagnetic code \textsc{calder} in 2D3V (2D in configuration space, 3D in momentum space)
geometry. The laser pulse, of wavelength $\lambda_0 = 1\,\rm \mu m$, is characterized by Gaussian spatial and temporal profiles of $5\,\rm \mu m$ FWHM spot size
and 30~fs FWHM duration, respectively. Its peak intensity is $I_0= 1.4 \times 10^{20}\,\rm W\,cm^{-2}$, corresponding to a dimensionless field strength $a_0 \simeq 10$.
Propagating in the $x>0$ direction  and linearly polarized along the $y$ axis, the laser pulse is focused at normal incidence onto fully ionized, sub-micron-thick
CH$_2$ foil targets with $1.3\,\rm g\,cm^{-3}$ density. These comprise three charged particle species (electrons, C$^{6+}$ and H$^+$ ions). The initial total
electron density is $n_{e0}=400\,n_c$. The targets are initialized with sharp gradients, a transverse width $D=300\,c/\omega_0 \simeq 47.7\,\rm \mu m$ and a
thickness varying in the range $15 \le d_0 \le 500 \,\rm nm$. The minimum target thickness considered ($d_0=15\,\rm nm$) is slightly above the predicted optimal
thickness, $d_{\rm opt} \simeq 12.5\,\rm nm$. Simulations at lower thicknesses were not carried out due to excessive numerical cost. 

The definition of the simulation mesh is a trade-off between having a domain large enough to allow propagation of the radiated fields over several 100~fs, and keeping the same spatio-temporal resolution and comparable data volume for all target sizes. The simulation domain has dimensions of $600 \times 600\,(c/\omega_0)^2 \simeq 95.5 \times 95.5\,\rm \mu m^2$. The grid size is $\Delta x = \Delta y = 0.03\,c/\omega_0 \simeq 4.8\,\rm nm$ (smaller than the target skin depth $c/\omega_p = 0.05c/\omega_0 \simeq 8\,\rm nm$) and the time step is $\Delta t=0.02\omega_0^{-1} \simeq 10\,\rm as$. The grid size is only one third of the thinnest (15~nm) target, but its ultrafast laser-driven expansion will lead to an interaction
effectively taking place over a broader, well resolved spatial region. Each particle species is initially represented by 400 macro-particles per cell
at $d_0= 500\,\rm nm$, and by 4000 macro-particles per cell at $d_0= 50\,\rm nm$ and $d_0=15\,\rm nm$ in order to achieve comparable statistics
in all cases. We did not investigate targets below the relativistic transparency threshold, because this would have forced us to reduce the spatial and time
steps by the same factor, hence increasing prohibitively the computational time.

\begin{figure}
  \centering
  \includegraphics[width=\columnwidth]{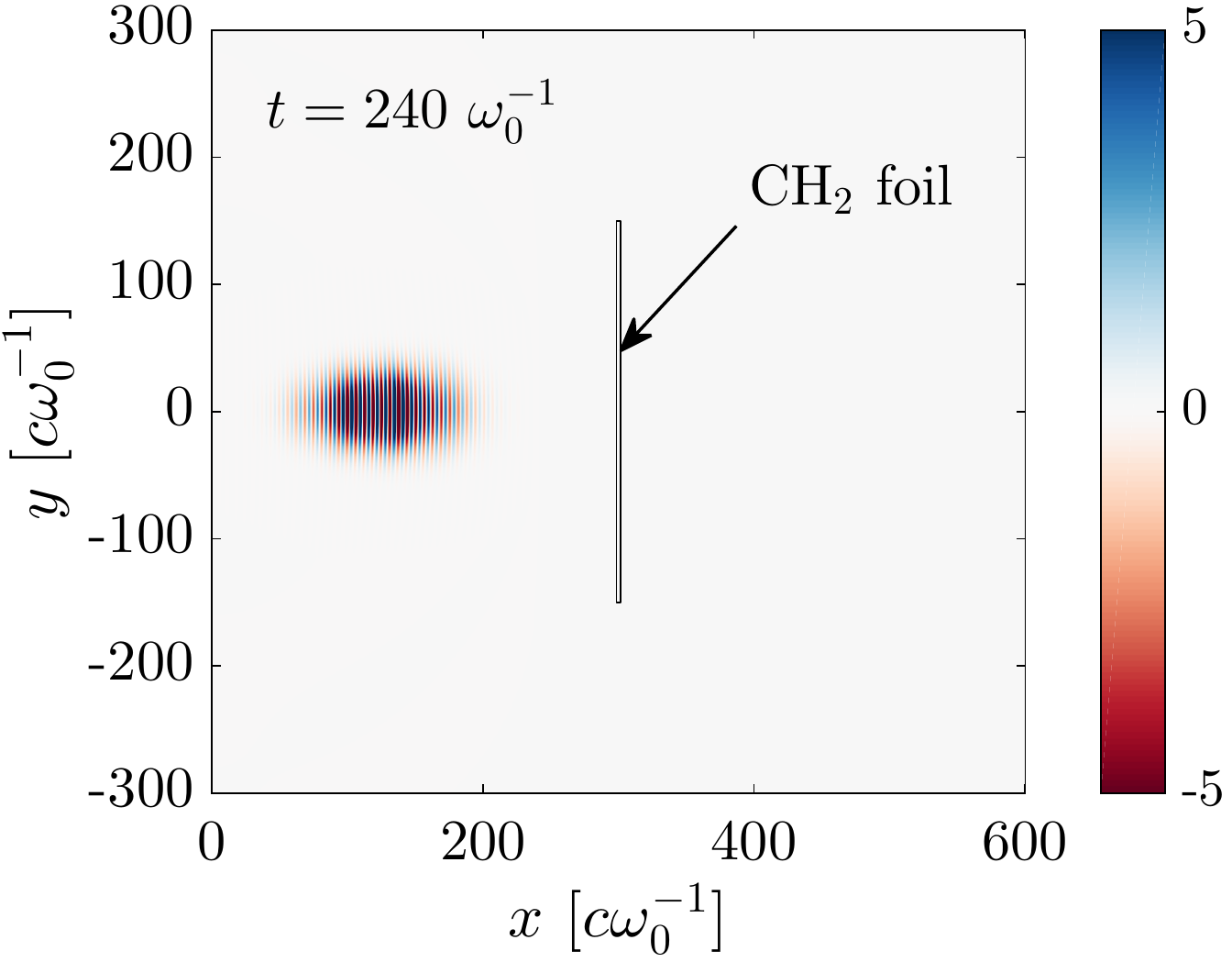}
  \caption{Snapshot of the incoming laser electric field $E_y$ [$m_e c \omega_0/e$] before its interaction with a 500~nm CH$_2$ foil target.}
  \label{fig:domain}
\end{figure}

Figure~\ref{fig:domain} illustrates the simulation setup before the pulse hits the target. In this figure, and as in the following ones, space and time are
normalized by $c/\omega_0 = 0.16\,\rm \mu m$ and $\omega_0^{-1} = 0.53\,\rm fs$.

\section{Main features of the laser-foil interaction}
\label{sec:main_features}

Before examining the THz radiation processes, let us first describe the electron and ion dynamics induced by the intense laser pulse in the $d_0=500\,\rm nm$ target. 

\begin{figure}
  \centering
 \includegraphics[width=\columnwidth]{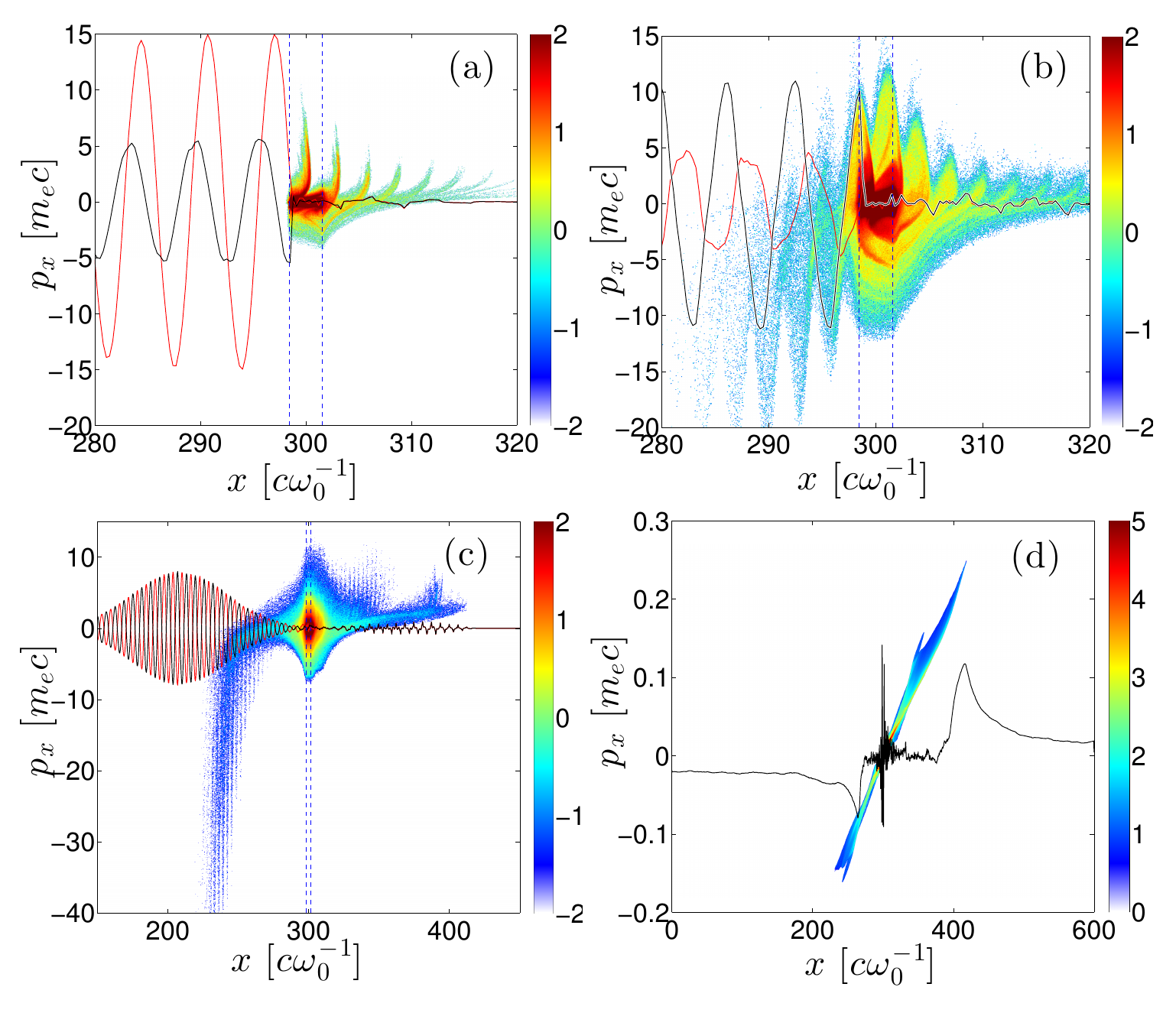}
  \caption{
 Overlap of the longitudinal electron phase space map ($x,p_x$) and of the on-axis $E_y$ (red curve) and $B_z$ (black curve) laser fields at (a) $t=400\,\omega_0^{-1}$, (b) $t=440\,\omega_0^{-1}$ and (c) $t=500\,\omega_0^{-1}$ for the 500-nm-thick target. (d) Proton longitudinal phase space $(x,p_x)$ and on-axis $E_x$ electrostatic field (averaged over one laser period) at $t = 1000\,\omega_0^{-1}$. The electric (resp. magnetic) fields are normalized to $m_e c \omega_0/e$ (resp. $m_e \omega_0/e$). Note the change of scale in (c,d) compared to (a,b). In (a,b,c) the vertical blue dashed lines indicate the initial target limits.}
\label{fig:phase_spaces}
\end{figure}

Figure~\ref{fig:phase_spaces} shows four successive snapshots of the longitudinal ($x,p_x$) electron phase space along with the on-axis laser fields ($E_y, B_z$).
The target initially extends over the region $298.44 \le x\omega_0/c \le 301.56$, and the laser pulse reaches its maximum at $t=420\,\omega_0^{-1}$.
At $t=400\,\omega_0^{-1}$ [Fig.~\ref{fig:phase_spaces}(a)], the laser's rising edge has hit the target. The highly overcritical electron density causes
strong laser reflection, and hence the formation of an electromagnetic standing wave in front of the target, with the $E_y$ and $B_z$ spatial extrema being separated by
$\lambda_0/4$ ($\pi/2$ in $c/\omega_0$ units). The two standing-wave patterns are also dephased in time by $\pi/2$ (in $\omega_0^{-1}$ units), which explains their nonequal
instantaneous amplitudes.

Owing to their small mass, the electrons react the fastest to the laser field, and are energized through a combination of vacuum and skin layer heating processes \cite{Bauer:pop:2007, May:pre:2011, Debayle:pop:2013}.  A fraction of them are accelerated forward by the laser's ponderomotive force in the form of $\lambda_0/2$-periodic jets, with maximum longitudinal momenta $p_x/m_ec \simeq a_0 \simeq 10$. As they flow into vacuum, a strong electrostatic sheath field ($E_x$) is induced which eventually reflects most of them back into the target and the (still present) laser wave (see the $p_x<0$ electrons at $x>301.6\,c/\omega_0$).
Upon interacting with the oscillating laser's ponderomotive force and the resulting charge-separation field, the recirculating  electrons develop an increasingly hot, symmetric $p_x$-distribution inside the target, which favors further laser-driven energization \cite{Bauer:pop:2007, May:pre:2011}. The energy spectrum of the forward-moving electrons is characterized by a decreasing exponential slope of $\sim 6.5\,\rm MeV$ (not shown). Some backward-moving electrons are energetic enough to overcome the ponderomotive potential and be injected into the standing wave [see Fig.~\ref{fig:phase_spaces}(b) recorded at $t=440\,\omega_0^{-1}$]. These can be re-accelerated to even larger negative momenta ($p_x \lesssim -40 m_e c$) by the reflected part of the laser pulse, a process known as vacuum laser acceleration \cite{Yu:prl:2000, Thevenet:np:2016} [see Fig.~\ref{fig:phase_spaces}(c) recorded at $t=500\,\omega_0^{-1}$]. 

The electrostatic sheath field ($E_x$) set up by the motion of the laser-driven electrons leads to ion acceleration \cite{Mora:prl:2003, *Mora:pre:2005, Brantov:prstab:2015}. The lighter protons rapidly separate from the carbon ions and reach the highest velocities. Figure~\ref{fig:phase_spaces}(d) shows the proton longitudinal phase space as recorded at $t= 1000\,\omega_0^{-1}$, overlaid with the longitudinal electrostatic field, averaged over a laser period. The forked shape of the phase space of $p_x>0$ protons originates from a combination of (prevailing) TNSA and radiation pressure acceleration. Some protons are also accelerated in the backward direction  $(p_x < 0$) as a result of TNSA at the target frontside. The electrostatic field structures associated with forward and backward TNSA are clearly seen at $x = 415\,c/\omega_0$ and $x = 263\,c/\omega_0$, respectively.

\begin{figure}
  \centering
  \includegraphics[width=\columnwidth]{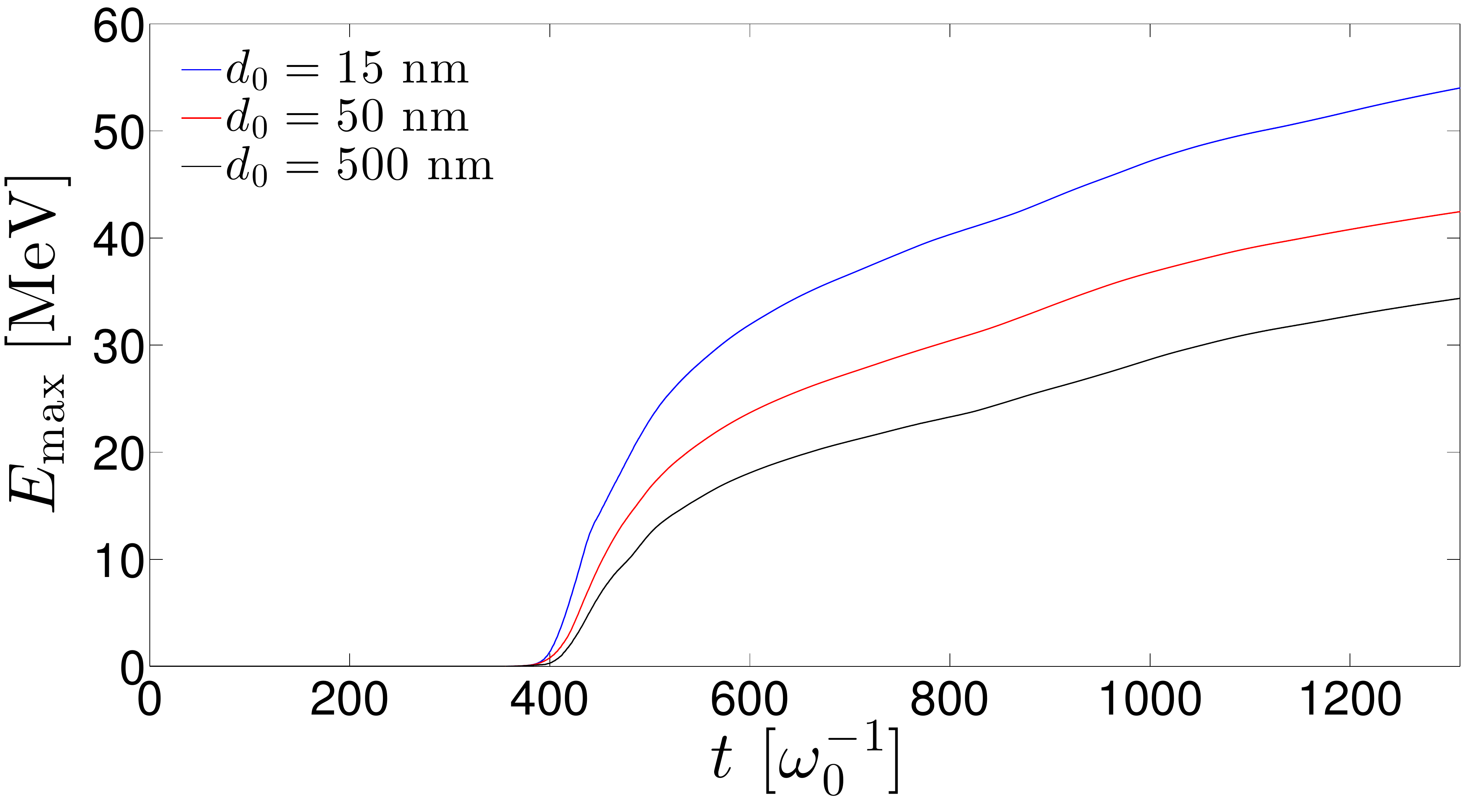}
  \caption{Maximum proton energy [MeV] over time [$\omega_0^{-1}$] for three target thicknesses (see legend).}
  \label{fig:nrj_max_H_vs_d}
\end{figure}

The black curve in Fig.~\ref{fig:nrj_max_H_vs_d} plots the temporal evolution of the instantaneous maximum energy, $E_{\rm max}(t)$, of the forward-moving protons ($v_x >0$). After a rapid growth at early times, $E_{\rm max}(t)$ increases more slowly by $t \simeq 500\,\omega_0^{-1}$, yet without showing a clear saturation trend up to simulation time $t=1300\,\omega_0^{-1}$. This sustained acceleration is a consequence of the reduced 2D geometry because of improper description of the momentum anisotropy and transverse dilution of the hot electrons \cite{Liu:pop:2013, Stark:pop:2017}. A faster drop in the sheath field strength is also expected in 3D once the ion front has travelled a distance comparable with the transverse size of the sheath, thus arresting
ion acceleration at lower kinetic energies \cite{Brantov:prstab:2015, Ferri:pop:2018}. 

The $E_{\rm max}(t)$ curves from simulations run with $d_0=50\,\rm nm$ (red) and $d_0=15\,\rm nm$ (blue) are also overlaid in Fig.~\ref{fig:nrj_max_H_vs_d}. The expected enhancement of the proton acceleration efficiency with decreasing foil thickness (down to the RSIT threshold) is manifest: at $t=1300\,\omega_0^{-1}$, $E_{\rm max}$ rises from $\sim 35\,\rm MeV$ at $d_0=500\,\rm nm$ to 54~MeV at $d_0=15\,\rm nm$. Further inspection of the numerical data reveals comparable absorbed fractions of the laser energy in the three simulation cases (viz. from $12.6\,\%$ at $d_0 =500\,\rm nm$ to $8.2\,\%$ at $d_0= 50\,\rm nm$).

\section{Analysis of the low-frequency radiation}
\label{sec:general_analysis}

\subsection{General picture of the THz field patterns}
\label{sec:overview}

We now study the low-frequency (i.e., in the THz domain) fields generated from the laser-plasma interaction. Spatial distributions of these fields $(\widetilde{E}_x,\widetilde{E}_y, \widetilde{B}_z)$ are obtained by Fourier transforming the original field distributions, applying a hypergaussian filter $\Pi(k) = \exp{[-(k/k_c)^{2n}]}$ and then inverse Fourier transforming. We here take $n = 6$ and a cut-off wavenumber $k_c = 0.3\,\omega_0/c$ associated with a frequency bandwidth $\nu_c = c k_c/2\pi = 90\,\rm THz$. 

\begin{figure}
  \centering
  \includegraphics[width=\columnwidth]{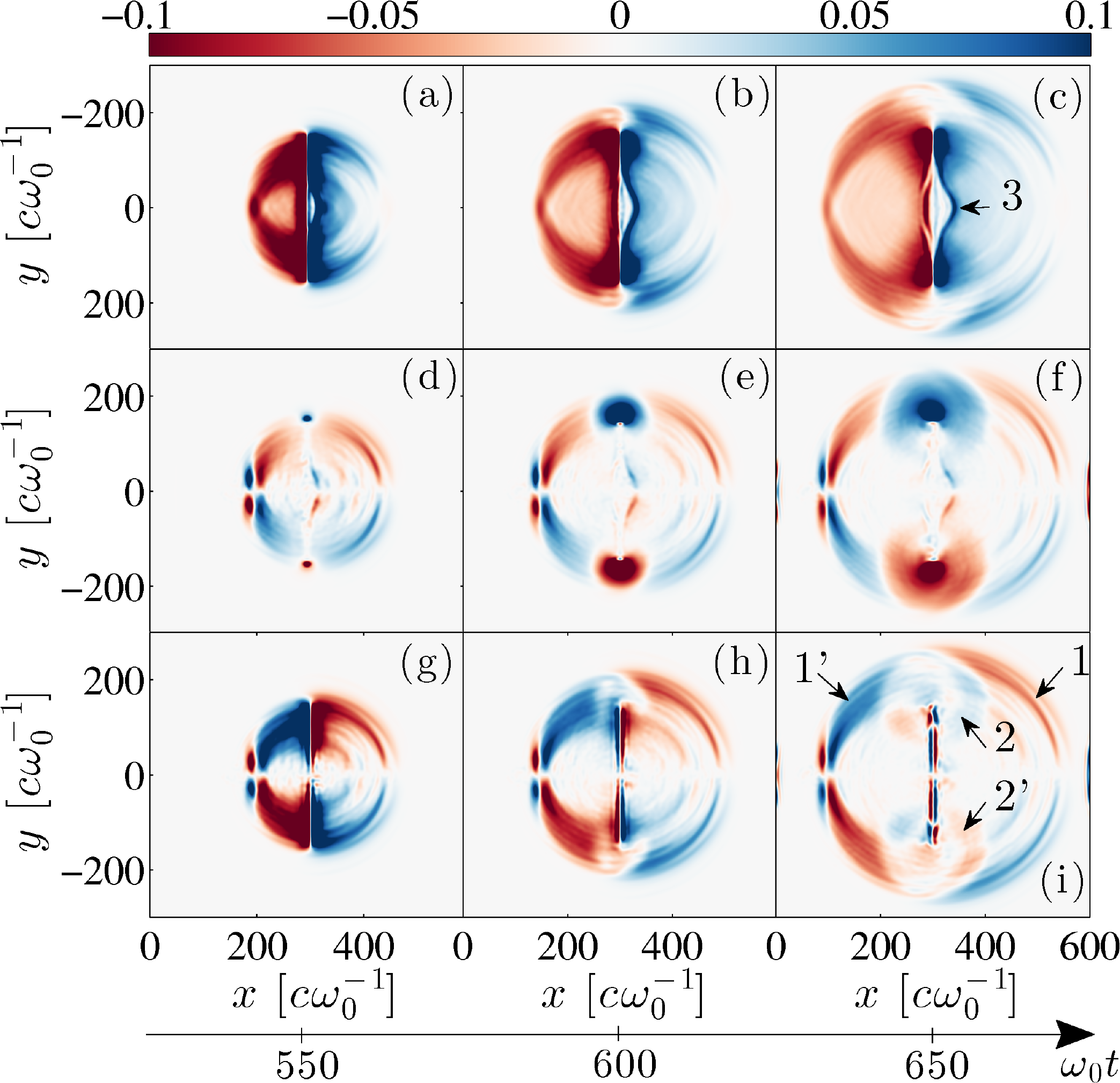}
  \caption{Spatial distributions of the low-frequency (a-c) $\widetilde{E}_x$ [$m_e c \omega_0/e$], (d-f) $\widetilde{E}_y$ [$m_e c \omega_0/e$] and (f-i) $\widetilde{B}_z$ [$m_e \omega_0/e$] fields at $t=550\,\omega_0^{-1}$, $t=600\,\omega_0^{-1}$ and $t=650\,\omega_0^{-1}$ (see time arrow) for the 500-nm-thick target. Those fields are filtered in the THz range ($\omega<0.3\,\omega_0$). The structures indicated by (1,1'), (2,2') and (3) are detailed in the text.}
  \label{fig:field_map}
  \end{figure}

An overview of their dynamics is provided by Figs.~\ref{fig:field_map}(a-i), which display the field distributions at three successive times following the laser interaction with the $500\,\rm nm$ target. Although the field maps exhibit rather complex patterns, one can discern the following main structures:
\begin{itemize}
\item Two outgoing radial waves originating from the laser spot and propagating in vacuum at the speed of light in the forward (1) and backward (1') directions, respectively. 
Those waves are visible in the three field maps. Their mutual spatial offset indicates that the backward wave has been emitted about $25\,\omega_0^{-1}$ after the
forward wave.
\item  Two outgoing  waves originating from the top (2) and bottom (2') edges of the foil, also propagating at the speed of light. Those waves have been synchronously emitted
$150\,\omega_0^{-1}$ after wave (1). The polarity of their $\widetilde{E}_y$ field [Figs.~\ref{fig:field_map}(d-f)] indicates that they are associated with an accumulation
of positive charges around the target edges. 
\item The electrostatic sheath field (3) at the boundary of the expanding protons. This field, which is responsible for TNSA, is most pronounced in the forward direction (i.e., the
preferential direction of the hot-electron flow at $d_0=500\,\rm nm$) and is mainly $x$-polarized. However, due to the transverse density gradient of the expanding proton cloud, it 
also comprises a $\widetilde{E}_y$ component, which is an odd function of the transverse coordinate ($y$).
\end{itemize}

\subsection{Identification of the radiating current sources} 
\label{sec:current_sources}

We now focus on the light-speed, low-frequency signals (1,1') and (2,2') revealed by the above snapshots of the $\widetilde{B}_z$ distribution. Note that in the present 2D3V simulation of a $p$-polarized laser pulse, the $\widetilde{B}_z$ field captures the full magnetic component of the low-frequency radiation. This contrasts with the $\widetilde{E}_x$ and  $\widetilde{E}_y$ components, which only carry a fraction of the radiated electric fields depending on their local polarization.

All radiated signals evidently result from the strong plasma currents generated by the laser-plasma interaction around the target. In particular, from previous works \cite{Ding:pre:2016,Liao:ppcf:2017, Herzer:njp:2018}, we expect the strong longitudinal currents carried by the hot electrons breaking off the target back (resp. front) side to emit coherent transition radiation in the forward (resp. backward) direction.  The transverse currents associated with the laterally moving hot electrons and
the shielding plasma electrons should also contribute to the overall radiation. It is therefore worthwhile to identify whether the observed radiated signals mostly result from the longitudinal ($j_x$) or transverse ($j_y$) component of the laser-driven plasma currents. 

To this purpose, we have added two Maxwell solvers in our code to advance separately the $(E_x,E_y,B_z)$  fields resulting from $j_x$ and $j_y$. Specifically, the $j_x$-driven fields $(E_x^\parallel,E_y^\parallel,B_z^\parallel)$ and $j_y$-driven fields $(E_x^\perp,E_y^\perp,B_z^\perp)$ are solved from the Maxwell-Faraday and Maxwell-Amp\`ere
equations:
\begin{align}
\frac{\partial E_x^\parallel}{c^2\partial t}
&=
\frac{\partial B_z^\parallel}{\partial y} - \mu_0 j_x \,,
\\
\frac{\partial E_y^\parallel}{c^2\partial t}
&=
- \frac{\partial B_z^\parallel}{\partial y} - 0 \,,
\\
\frac{\partial B_z^\parallel}{\partial t}
&=
\frac{\partial E_x^\parallel}{\partial y} - \frac{\partial E_y^\parallel}{\partial x} \,,
\\
\frac{\partial E_x^\perp}{c^2 \partial t}
&=
\frac{\partial B_z^\perp}{\partial y} - 0 \,,
\\
\frac{\partial E_y^\perp}{c^2 \partial t}
&=
- \frac{\partial B_z^\perp}{\partial y} - \mu_0 j_y \,,
\\
\frac{\partial B_z^\perp}{\partial t}
&=
\frac{\partial E_x^\perp}{\partial y} - \frac{\partial E_y^\perp}{\partial x} \,.
\end{align}
The resulting fields are then filtered as explained above.

\begin{figure}
  \centering
  \includegraphics[width=\columnwidth]{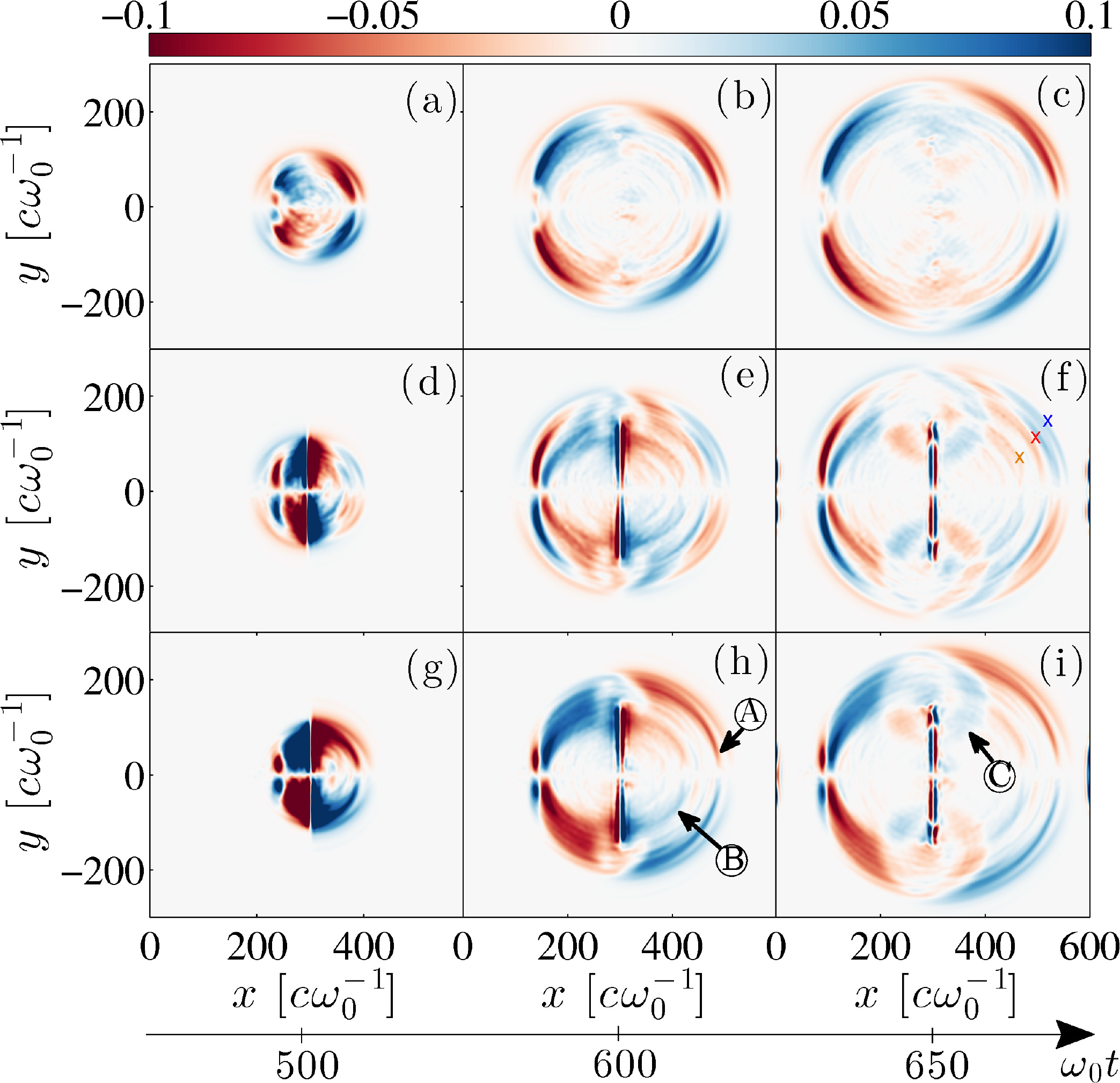}
  \caption{
  Spatial distributions of (a,b,c) $\widetilde{B}_z^\parallel$, (d,e,f) $\widetilde{B}_z^\perp$ and (g,h,i) $\widetilde{B}_z$ [$m_e \omega_0/e$] fields filtered in the THz
  range ($\omega<0.3\omega_0$) at $t=500\,\omega_0^{-1}$, $t=600\,\omega_0^{-1}$ and $t=650\,\omega_0^{-1}$ (see time arrow) for the 500-nm thick target. 
  The blue, red and brown cross symbols in (f) pinpoint the radiated field structures discussed in Sec.~\ref{subsubsec:transverse_dynamics}.}
\label{fig:decoupling_bz}
\end{figure}

\subsubsection{Longitudinal electron dynamics}

Figures~\ref{fig:decoupling_bz}(a-i) present the results of this procedure at three different instants (i.e., $t=500\,\omega_0^{-1}$, $600\,\omega_0^{-1}$ and $650\,\omega_0^{-1}$). The first row shows $\widetilde{B}_z^\parallel (t)$, the second one $\widetilde{B}_z^\perp (t)$ and the third one the total filtered field $\widetilde{B}_z (t) = \widetilde{B}_z^\parallel (t) + \widetilde{B}_z^\perp (t)$. Note that Figs.~\ref{fig:decoupling_bz}(h-i) reproduce Figs.~\ref{fig:field_map}(h-i) to help the reader connect the radiation patterns discussed in the following to the main low-frequency structures presented in the previous subsection. 

Figures~\ref{fig:decoupling_bz}(a-c) demonstrate that, as expected, the two radial waves (1) and (1') outgoing from both target sides are induced by $j_x$. We ascribe these waves to CTR by the longitudinal hot-electron motion across the target surfaces. Accordingly, their polarity fulfills $\mbox{sgn} (B_z ({\vec r})) = \mbox{sgn}( {\vec j}_s \times ({\vec r} - {\vec r}_s))$, with ${\vec j}_s$ the hot-electron current source at the target center ${\vec r}_s$. Their relative $\sim 25\,\omega_0^{-1}$ delay corresponds to the reflection time of the initially forward-moving hot electrons by the sheath field set up at the target boundaries [see $\widetilde{E}_x$ in Figs.~\ref{fig:field_map}(a-c)].  When viewed along a fixed direction from the target center, the two waves take on the form of half-cycle pulses with similar field strength as a function of the distance from the laser spot. At $t=650\,\omega_0^{-1}$ (i.e., $230\,\omega_0^{-1}$ after the on-target laser peak), one finds
$\widetilde{B}_z \simeq 0.1\,m_e \omega_0/e$, corresponding to an electric field $\vert \widetilde{\mathbf{E}} \vert = c B_z \simeq 3\times 10^{11}\,\rm V\,m^{-1}$. The subsequent longitudinal motion of the plasma leads to much weaker (by more than an order of magnitude) radiated field strengths.

Figures~\ref{fig:decoupling_bz}(a-c) also disclose a backward-propagating dipolar (as seen along $y$) $\widetilde{B}_z^\parallel$ structure (located at $x \simeq 230\,c\omega_0$ at $t=500\,\omega_0^{-1}$). This structure corresponds to the inner self-field of the relativistic electron bunch accompanying the reflected laser pulse. This bunch has a density minimum on axis due to the transverse ponderomotive force resulting from the laser intensity gradients. The region of maximal current density is delineated by the curve of vanishing $\widetilde{B}_z^\parallel$ field surrounding the blue and red dots in Fig.~\ref{fig:decoupling_bz}(a-c). Since the bunch's speed is very near that of light, the outer portion of its self-field is merged with the backward-emitted CTR field \cite{Carron:pier:2000}.

\subsubsection{Transverse electron and early-ion dynamics}
\label{subsubsec:transverse_dynamics}

While the radial waves (1) and (1') radiated from the target center are mainly driven by $j_x$, one can see from Figs.~\ref{fig:decoupling_bz}(d-f) that they also include a contribution from $j_y$. Unlike the essentially single-pulse shape of $\widetilde{B}_z^\parallel$, the $\widetilde{B}_z^\perp$ radiation shows a multi-pulse profile. In the forward direction, the primary $\widetilde{B}_z^\perp$ pulse [indicated by a blue cross in Fig.~\ref{fig:decoupling_bz}(f)] has a polarity ($=\mathrm{sgn} (y)$) opposite to that of the coincident $\widetilde{B}_z^\parallel$ pulse, which corresponds to a rotation by $\pi/2$ of the coincident $\widetilde{B}_z^\parallel$ field. We attribute this radiation to the transverse (along $\pm y$) motion of the diverging hot electrons crossing the target backside. 

Two $\widetilde{B}_z^\perp$ pulses are subsequently emitted, separated by  $\sim 35\,c/\omega_0$ and with an opposite polarity ($=-\mathrm{sgn}(y)$) [red cross in Fig.~\ref{fig:decoupling_bz}(f)]. We interpret them as resulting from the shielding surface currents induced when the hot electrons exit the target. These electron currents act both to screen the radial waves
and the static fields generated along the target surfaces by the outward expanding hot electrons. In terms of radiation, they are equivalent to outward pulses of positive charge propagating at (almost) the velocity of light within a skin-depth layer. The radiation they produce is therefore analogous to that of a pulse-excited line antenna \cite{Smith:ajp:2001} (with the caveat, however, that in a realistic 3D geometry, the target we simulate here would not correspond to a wire antenna but to a conducting plate of infinite $z$-extent). 

The surface return current pulses first emit a radial wave (located at $x\simeq 505\,c/\omega_0$ at $t=650\,\omega_0^{-1}$) when triggered at the target center [Fig.~\ref{fig:decoupling_bz}(f)]. This is similar to radiation by two positively charged bunches suddenly accelerated in the $\pm y$ directions. The origin of the second radial wave [of same polarity and located at
$x\simeq 470\,c/\omega_0$ at $t=650\,\omega_0^{-1}$, see the brown cross in Fig.~\ref{fig:decoupling_bz}(f)] appears to be correlated with the subsequent deformation of the target surface, entailing current perturbations around the laser spot. This local deformation, caused by the early ion expansion, accounts for the dipolar $\widetilde{E}_y$ field profile observed
around $x=300\,c/\omega_0$ in Figs.~\ref{fig:field_map}(d-f).

In the backward direction, the $\widetilde{B}_z^\perp$ distribution exhibits mainly two radial pulses of opposite polarity, similar to the two primary forward pulses. Interestingly, the backward pulses appear to be about twice more intense than the equivalent forward pulses, due to the larger transverse current carried by the backward-moving electron bunch. 

\begin{figure}
  \centering 
\includegraphics[width=0.75\columnwidth]{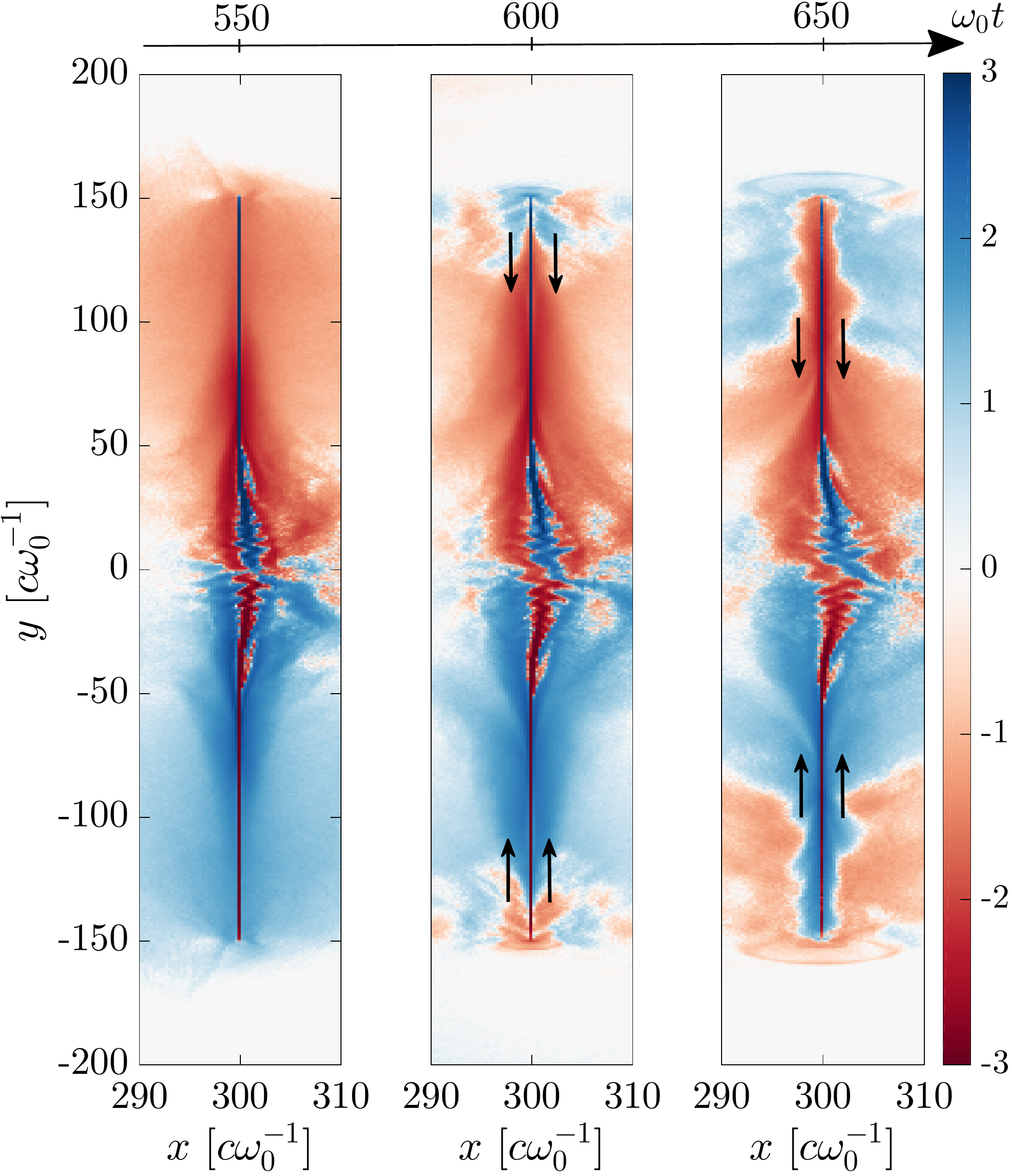}
  \caption{Snapshots (see time arrow) of the transverse current $j_y$ [$e c n_c$] reaching the targets ends at $d_0=50\,\rm nm$. The displayed quantity is 
  $\mathrm{sgn}(j_y)\left\{ 3 + \log_{10} [\mathrm{max} (10^{-3},\vert j_y \vert )] \right\}$ so as to visualize positive and negative $j_y$ values in a log-like fashion. 
  Black arrows show the motion of the transversely refluxing hot electrons. The dynamics of surface currents is not visible due to the limited graphical resolution.
  }
  \label{fig:edge_current}
\end{figure}

Again, similarly to a wire antenna, radial waves are emitted when the shielding surface currents ($j_y$) reach, and reflect at, the upper and lower ends of the foil. These waves obviously correspond to waves (2) and (2') as defined in Fig.~\ref{fig:field_map}(i). This process was analyzed in Ref.~[\onlinecite{Zhuo:pre:2017}] using the theory of thin-wire antennas \cite{Smith:ajp:2001}. The arrival of the shielding currents at the terminations also leads to local accumulations of positive charge, resulting in the strong Coulomb $\widetilde{E}_x$ and $\widetilde{E}_y$ fields seen in Figs.~\ref{fig:field_map}(d-f). These fields tend to decelerate and reflect the hot electrons passing through the target ends, as evidenced by the successive snapshots of $j_y$ of Fig.~\ref{fig:edge_current}. The decelerated/reflected hot electrons should also produce some radiation, but with opposite polarity to that by the shielding currents. Overall, though, we find that the latter are mainly responsible for the radiation from the target ends. The inward reflection of the hot electrons is accompanied by the reflection of the surface currents (barely visible in Fig.~\ref{fig:edge_current} due to limited resolution). This explains why the waves radiated from the target ends are not directed in the outward direction, but instead appear to be essentially isotropic \cite{Smith:ajp:2001}. Another consequence of the reversal of the hot-electron and surface currents  is the formation of inward-moving magnetic-field nodes along the target, as observed in Figs.~\ref{fig:decoupling_bz}(f,i).

In summary, we have identified several THz signals radiated during the first $\sim 100\,\rm fs$ following the laser interaction $(t \le 650\,\omega_0^{-1}$). First, two CTR-type bursts [radial waves (1) and (1') in Fig.~\ref{fig:field_map}(i)], lasting a few $10\,\omega_0^{-1}$, are emitted from the irradiated region, in both the forward and backward directions. They are associated with the hot electrons crossing first the backside and then, due to electrostatic reflection, the front side of the target. The longitudinal component of the hot-electron current is the main source of those waves. Subsequent emission of weaker THz waves from the laser spot is ascribed to the shielding surface electron currents accompanied by early foil deformations.  The arrival of these surface currents at the target ends (after a delay of $D/2c$) results in the antenna-like emission of two other radial waves [(2) and (2') in Fig.~\ref{fig:field_map}(i)].

\section{Influence of the foil thickness on the low-frequency radiation}
\label{sec: foil_thickness}

\begin{figure}
  \centering 
  \includegraphics[width=\columnwidth]{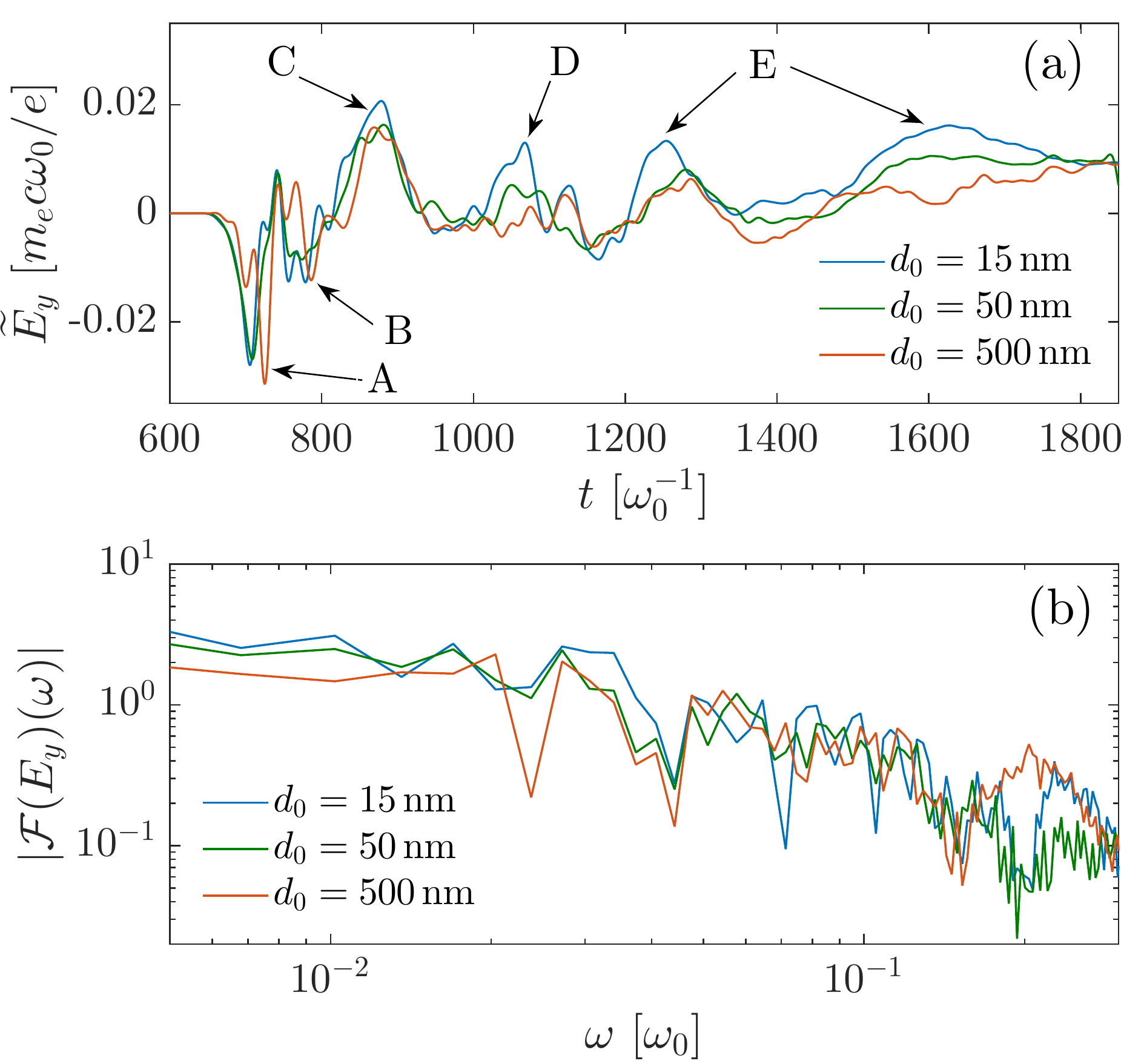}
  \caption{(a) Waveform of the low-frequency electric field $\widetilde{E}_y$ [$m_e c\omega_0/e$] and (b) corresponding spectra as recorded at
  location ($x,y) = (600,100)\,c/\omega_0$ for the three target thicknesses (see legend). Circles (A), (B), (C), (D) and (E) in (a) indicate the successive
  signals discussed in the text.}
  \label{fig:probe_ey}
\end{figure}

The simulations run with 15~nm and 50~nm-thick foil targets exhibit qualitatively similar features to those observed at $d_0=500\,\rm nm$.  To assess quantitatively the dependence of the low-frequency radiation on the target thickness, we have recorded the $\widetilde{E}_y$ field as a function of time at the fixed location $(x,y) = (600, 100)\,c/\omega_0$. The temporal waveforms of $\widetilde{E}_y$ obtained at $d_0=15\,\rm nm$, $50\,\rm nm$ and $500\,\rm nm$ and their corresponding spectra are plotted in Figs.~\ref{fig:probe_ey}(a) and~\ref{fig:probe_ey}(b), respectively. The three temporal profiles show a similar sequence of signals, albeit
with differences in magnitude and fine-scale modulations.

The first electric-field burst (A), of negative polarity, corresponds to the arrival of wave (1) at the detector [see also Figs.~\ref{fig:field_map}(e) and \ref{fig:decoupling_bz}(h)]. Its time of arrival ($t\simeq 700-720\,\omega_0^{-1}$) is consistent with the time-of-flight of wave (1) emitted from the target center ($x=300\,c/\omega_0$) at $t \simeq 400\,\omega_0^{-1}$. The signals produced by the two thinnest targets are almost identical, with an amplitude of $\sim 0.028\,m_e c \omega_0/ e \simeq 84\,\rm GV\,m^{-1}$ and a pulse duration of $\sim 25\,\omega_0^{-1} \simeq 13\,\rm fs$.  At $d_0=500\,\rm nm$, this signal features a prepulse (also visible in Figs.~\ref{fig:field_map} and \ref{fig:decoupling_bz}). Its maximum is delayed by  $\sim 20\,\omega_0^{-1}$, as a result of temporal modulations in the hot-electron source induced by the early interaction of the laser pulse. 

The second signal (B), of mainly negative polarity, is detected at $t \simeq 750-850\,\omega_0^{-1}$. It corresponds to the weaker secondary radial wave emitted from the irradiated region as seen in Fig.~\ref{fig:decoupling_bz}(h), which we attributed to local perturbations in the shielding $j_y$ current. 

The third radiation burst (C), of positive polarity, is related to the radial wave (2) from the upper target end. Its detection time ($t \simeq 850\,\omega_0^{-1}$) agrees with an emission starting at $t \simeq 550\,\omega_0$, i.e., $D/2c$ later than wave (1). Similar field strengths ($\widetilde{E}_y \simeq 0.02\,m_e c\omega_0/e \simeq 60\,\rm GV\,m^{-1}$) are recorded for the three foil thicknesses.

The fourth signal (D), observed at $t \simeq 1050\,\omega_0^{-1}$, is linked to the lateral recirculation of the hot electrons and of the shielding surface currents. When the shielding currents flowing along the target backside return to the center, they encounter a bended surface due to the target expansion. The rightward (towards vacuum) acceleration that they then experience results in additional radiation (with positive polarity). Since the target deformation is enhanced with decreasing foil thickness, we expect this radiation to be maximized at $d_0=15\,\rm nm$, as is indeed observed in Fig.~\ref{fig:probe_ey}. 

\begin{figure}
  \centering
\includegraphics[width=\columnwidth]{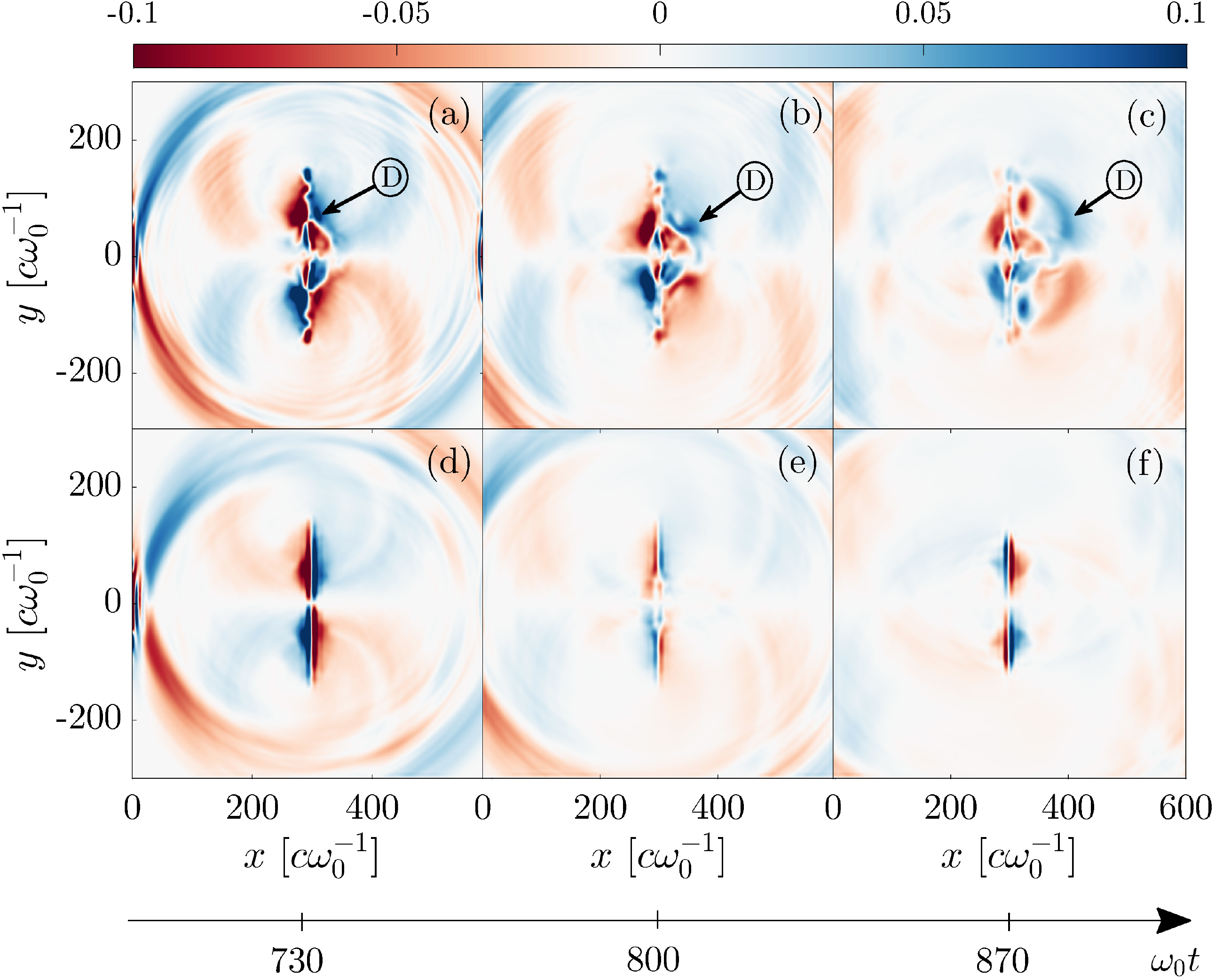}
  \caption{Snapshots (see time arrow) of the low-frequency magnetic field $\widetilde{B}_z$ [$m_e \omega_0/e$] with (a,b,c) mobile and (d,e,f) fixed ions for the 15~nm thick foil target. Circle (D) shows the radiation from the deflected shielding currents.}
  \label{fig:deflection}
\end{figure}

To further support this scenario, we have re-run the $d_0=15\,\rm nm$ simulation with fixed ions. The idea is that immobile ions should prevent target deformations and,therefore, the related emission. Such behavior is clearly demonstrated in Fig. \ref{fig:deflection}, which displays successive snapshots of $\widetilde{B}_z$ with either mobile (top row) or immobile  (bottom row) ions. The two maps at $t = 730\,\omega_0^{-1}$ show the CTR-type emission from the target center and the edge emission
from the shielding currents. The latter turns out to be weaker with immobile ions, owing to a lower laser-to-hot-electron conversion efficiency, and hence weaker shielding currents. With mobile ions, the quasistatic magnetic-field structures formed around the target center clearly reveal the local deformations of the target boundaries [see Figs.~\ref{fig:deflection}(a,b,c)]. At $t=800\,\omega_0^{-1}$, the transversely refluxing electron currents reach the deformed central part of the target where they are deflected rightward. The resulting emission is seen to emerge from the target backside at $t=870\,\omega_0^{-1}$ [D in Fig.~\ref{fig:deflection}(c)]. This radiation will be recorded $\sim 200\,\omega_0^{-1}$ later at the probe location. With fixed ions, by contrast, the converging electron currents smoothly stream though each other [Figs.~\ref{fig:deflection}(d,e,f)], and so no radiation ensues. 

The above radiation process is also operative at $d_0= 50\,\rm nm$ (although weaker than at $d_0= 15\,\rm nm$ because of slower target expansion), but is found to be insignificant at $d_0=500\,\rm nm$. 

Our final comments will address the late-time ($t>1200\,\omega_0^{-1}$) THz signals indicated by label (E) in Fig.~\ref{fig:probe_ey}(a). This figure shows that these signals, of typical duration $\sim\,200\,\omega_0^{-1}$, are enhanced in thinner targets. This translates into a stronger spectral amplitude at 15~nm around $\omega_0 \simeq 0.03\,\omega_0$ as seen in Fig.~\ref{fig:probe_ey}(b). The three broadband spectra show the same decreasing trend at higher frequencies, up to minor variations near the THz cutoff value $\omega_c = 0.3\,\omega_0$.  

\begin{figure}
  \centering 
\includegraphics[width=\columnwidth]{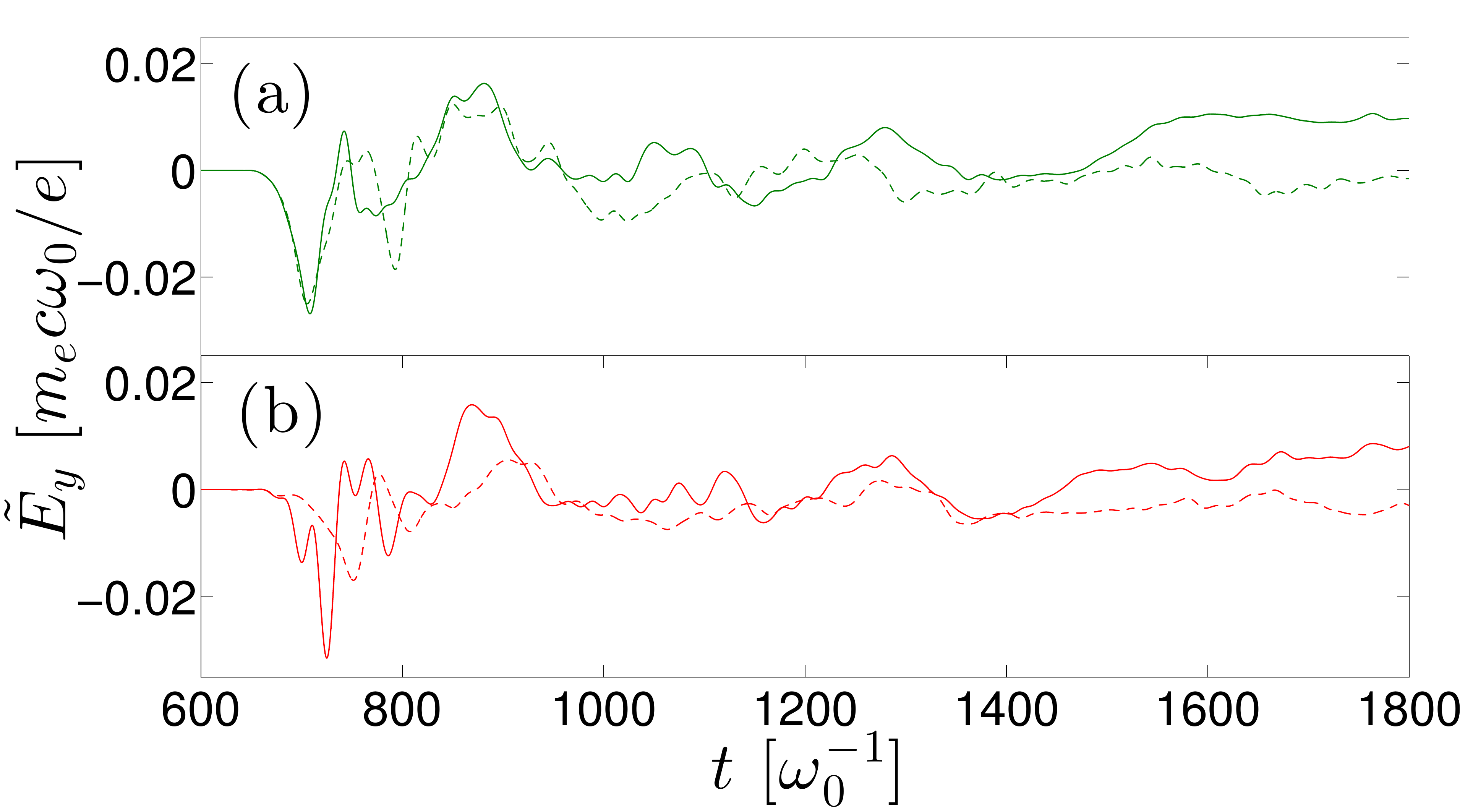}
  \caption{Waveform of the low-frequency electric field $\widetilde{E}_y$ [$m_e c \omega_0/e$] as recorded at location $(x,y)=(600,100)\,c\omega_0^{-1}$ for (a) $d_0 =50\,\rm nm$ and (b) $d_0=500\,\rm nm$ with mobile (solid lines) and immobile (dashed lines) ions.}
  \label{fig:probe_ey_50_500}
\end{figure}

Moreover, the THz temporal signals are strengthened when allowing for mobile ions, as illustrated by Fig.~\ref{fig:probe_ey_50_500} in the $d_0=50\,\rm nm$ and $d_0=500\,\rm nm$ cases. Similar properties would be expected from the sheath-induced radiation ~\cite{Gopal:njp:2012, *Gopal:prl:2013, *Gopal:ol:2013, Herzer:njp:2018}. In particular, the target thickness dependency (e. g., the late-time emission of multiple pulses amplified in thinner targets) qualitatively agrees with the experimental findings of Jin \emph{et al.}~\cite{Jin:pre:2016} recalled in our introduction. However, we have not been able to provide unambiguous evidence for the sheath radiation mechanism, due to the difficulty of discriminating between the radiated fields and the quasistatic fields attached to the particles (electrons and ions) that have then reached the probe.

\section{Conclusion}
\label{sec:conclusion}

By means of 2D PIC simulations, we have studied numerically the various processes of THz radiation from the interaction of an ultraintense femtosecond laser pulse with submicron-thick foil targets. The complex dynamics of the laser-driven hot electrons and associated surface shielding currents leads to the emission of several successive bursts. Two main types of radiation have been identified: CTR-type waves generated from the irradiated region by the forward-accelerated, and subsequently reflected, hot electrons, and antenna-type waves produced when the shielding surface currents leave the irradiated region, reflect off the target edges, or are deflected sideways along the deformed surface of the expanding plasma bulk. These secondary radiations appear to be maximized in thin targets close to the RSIT threshold, which expand, and hence deform, the fastest. Our analysis has been carried out by resolving, for the first time, the respective contributions of the longitudinal and transverse plasma currents to the low-frequency radiation. We believe that our mapping of the different source terms will help interpret the future experiments in this field.

\section*{Author's Contributions}
All authors contributed equally to this work.

\section*{Acknowledgments}
We acknowledge the ``Grand Equipement de Calcul Intensif'' (GENCI) for granting us access to the supercomputer IRENE under the grants No. A0070506129 and No. A0080507594.

\section*{Data availability}
The data that support the findings of this study are available within the present article. Complementary data can be made available from the corresponding authors upon reasonable request.

\bibliography{references} 

\end{document}